\documentclass[12pt]{iopart}
\usepackage{epsfig}

\def\dddot#1{\mathinner{\buildrel\vbox{\kern5pt\hbox{...}}\over{#1}}}
\def\n{\nonumber}

\def\ve{{\varepsilon}}

\def\d{{\rm d}}
\def\a{\alpha}
\def\b{\beta}
\def\bs{\begin{subequations}}
\def\es{\end{subequations}}
\def\be{\begin{equation}}
\def\ee{\end{equation}}
\def\bq{\begin{eqnarray}}
\def\eq{\end{eqnarray}}
\def\beq{\begin{eqnarray*}}
\def\eeq{\end{eqnarray*}}
\def\ba{\begin{eqnarray}}
\def\ea{\end{eqnarray}}
\def\p{\partial}
\begin{document}

\title{Gravitating fluids with Lie symmetries}

\author{ A M Msomi\dag\
 \footnote[2]{ Permanent address: Department of Mathematical Sciences,
 Mangosuthu University
of Technology,  P. O. Box 12363, Jacobs  4026, South Africa},
  K S Govinder\dag  and S D Maharaj\dag}
\address{\dag\ Astrophysics and Cosmology Research Unit,
 School of Mathematical Sciences,
 University of KwaZulu-Natal, Private Bag X54001,
 Durban 4000, South Africa \\ Email: maharaj@ukzn.ac.za
 }

\begin{abstract}
We analyse the underlying nonlinear partial differential equation
which arises in the study of gravitating flat fluid plates of
embedding class one. Our interest in this equation lies in
discussing new solutions that can be found by means of Lie point
symmetries. The method utilised reduces the partial differential
equation to an ordinary differential equation according to the Lie
symmetry admitted. We show that a class of solutions found
previously can be characterised by a particular Lie generator.
Several new families of solutions are found explicitly. In
particular we find the relevant ordinary differential equation for
all one-dimensional optimal subgroups; in several cases the ordinary
differential equation can be solved in general. We are in a position
to characterise particular solutions with a
 linear barotropic equation of state.\end{abstract}


\maketitle

\section{Introduction}
The local isometric embedding of four-dimensional Riemannian
manifolds $M_{4}$ in higher dimensional flat pseudo-Euclidean spaces
$E_{N}(N\leq 10)$ is important for several applications in general
relativity. For the basic theory and general results pertinent to
embeddings the reader is referred to Stephani {\it et al} (2003).
The invariance of the embedding class naturally generates a
classification scheme for all solutions of the field equations in
terms of their embedding class. The embedding class $p$ is the
minimum number of extra dimensions of the Riemannian manifold
$M_{4}$, ie. $p = N-4$. Exact solutions have been found by the
method of embedding in particular spacetimes for simple cases of low
embedding class. Some of these exact solutions may not be easily
found using other methods and techniques. For example the embedding
method has been utilised to find all conformally flat perfect fluid
solutions, in embedding class $p=1$, of Einstein's field equations
(Krasinki 1997, Stephani 1967a, Stephani 1967b). We point out that
embedding of four-dimensional Riemannian manifolds in higher
dimensional spacetimes with arbitrary Ricci tensors has been
investigated by several authors. The physical motivation here is to
understand the nature of physics in higher dimensions; the modern
view is that the Riemannian manifold $M_{4}$ is a hypersurface in
the
 higher dimensional bulk in the brane world scenario and other higher dimensional themes
  (Dahia and Romero 2002a, Dahia and Romero 2002b, Dahia {\it et al} 2008).

Gupta and Sharma (1996) have generated a relativistic model in higher dimensions  describing
gravitating fluid plates. Advantages of this model are that it is easy to interpret the physical features
using embedding in higher dimensions and the underlying differential equation governing the
gravitational dynamics is tractable. This model is expanding and not conformally flat.
A plane symmetric metric in four-dimensional spacetimes $M_{4}$ given by
\begin{equation}
\d s^2 = -\d r^2-t^2(\d\theta^2+\theta^2\d\phi^2)+(1+2\dot{V})\d
t^2+2V'\d r \d t \label{dad}
\end{equation}
is embedded in the five-dimensional pseudo-Euclidean space $E_{5}$ with metric
\begin{equation}
\fl\d s^2 = -(\d z^1)^2-(\d z^2)^2-(\d z^3)^2+(\d z^4)^2-(\d
z^5)^2. \label{dads}
\end{equation}
This embedding is achieved by setting
\numparts
\begin{eqnarray} z^1 &=& t\theta \cos \phi\\
z^2 &=& t\theta \sin \phi \\
z^3 &=& \frac{\theta^2}{2}t+V\\
z^4 &=& t\left(\frac{\theta^2}{2}+1 \right)+V\\
z^5 &=& r
\end{eqnarray}
\endnumparts
where $V = V(r,t)$ is an arbitrary function.

Consequently this model has embedding class $p = 1$ which allows
both conformally flat and nonconformally
 flat fluid distributions. The solutions admitted may be geodesic or accelerating. For a nonzero conformal (Weyl)
  tensor a partial differential equation has to be satisfied. This pivotal equation governs the evolution of the
  system and a particular class of solutions was identified by Gupta and Sharma (1996) by inspection.
 A detailed analysis of the pivotal equation shows that other classes of solution
 are possible which contain the Gupta and Sharma (1996) models as a special case.

Our intention is to systematically study the pivotal equation and to obtain a deeper insight into the nature
of solutions permitted using the Lie analysis of differential equations. In $\S2$ we discuss the fundamental
partial differential equation that governs the gravitational behaviour of the model, and present known solutions.
An outline of the basic features of the Lie symmetry analysis is given in $\S3$. We regain the Gupta and Sharma (1996)
 models using the relevant Lie generator in $\S4$. In $\S5$ we systematically study the group invariant solutions
 admitted by the fundamental equation. The partial differential equation is reduced to an ordinary differential
 equation for each element of the optimal system. The integrability of the ordinary differential equation is
 considered in each case and exact solutions are identified. The physical features of the model are
 discussed in $\S6$ and equations of state are found for particular solutions. Some brief concluding comments are made in $\S7$.

\section{The model}

The embedding of the four-dimensional Riemannian metric (\ref{dad})
into the five-dimensional flat metric (\ref{dads}) leads to a
differential equation that is central to the model. Gupta and Sharma
(1996) show that the pivotal equation is \ba \fl
 -V_{tt}V_{rr}+V_{rt}{}^{2} + \frac{1}{t}
\left[(1+2V_{t})V_{rr}-V_{tt}-2V_{rt}V_{r}\right] \n \\
+\frac{1}{t^2}\left[1+2V_{t}+V_{r}{}^{2}\right] = 0 \label{don}
 \ea
where subscripts denotes partial differentiation. This is a nonlinear equation in $V$ and difficult to solve. We need to explicitly
solve (\ref{don}) to describe the gravitational dynamics.

To demonstrate a class of solutions to (\ref{don}), Gupta and Sharma (1996) made the following assumption
\begin{equation}
 V = C \left(\frac{f(r)}{t} + h(t) \right) + C_1 \label{1a}
 \end{equation}
where $C$ and $C_1$ are arbitrary constants. Then (\ref{don}) reduces to the separable form
\begin{equation}
 4 f - \frac{4 C f_{r}{}^2}{C f_{rr}+1} = t^2 \left[\frac{1}{C} - th_{tt} + 2 h_{t}\right] = \a \label{2}
 \end{equation}
where $\a $ is the constant of separability. It is possible to solve the equation (\ref{2}) in terms of $t$ explicitly as
\begin{equation}
 h = - \frac{\a}{4 t} - \frac{t}{2C} - \frac{\a_1 t^3}{3} + \a_2 \label{3}
 \end{equation}
and to provide four solutions to the equation in terms of $r$, {\it viz}
\numparts
\begin{eqnarray}
f &=& \frac{1}{2 m^2 C} \sin X + \frac{1}{2m^2C} + \frac{\a}{4} \label{4a} \\
f &=& \frac{\a}{4} \\
f &=& \frac{1}{2C} (r+\b)^2 + \frac{\a}{4} \\
f &=& \frac{1}{2 m^2 C} \cosh X - \frac{1}{2m^2C} + \frac{\a}{4} \label{4d}
\end{eqnarray}
\endnumparts
where we have set $X=\sqrt{2}mr+m_0$ and $\a_1, \a_2, \b, m$ and $m_0$ are arbitrary constants.\\

Equations (\ref{3}) and (\ref{4a})--(\ref{4d}) are then combined to provide solutions to the original equation (\ref{don}), {\it viz}
\numparts
\begin{eqnarray}
V &=& \frac{1}{2 m^2 t} \sin X + \frac{1}{2m^2t}  - \frac{t}{2} - \frac{k t^3}{3} + k_1  \label{5a} \\
V &=& - \frac{t}{2} - \frac{k t^3}{3} + k_1\\
V &=& \frac{1}{2t} (r+\b)^2  - \frac{t}{2} - \frac{k t^3}{3} + k_1\\
V &=& \frac{1}{2 m^2 t} \cosh X - \frac{1}{2m^2t}
- \frac{t}{2} - \frac{k t^3}{3} + k_1
\label{5d}
\end{eqnarray}
\endnumparts
where $k=C\a_1$ and $k_1=C\a_2$.  Thus the assumption (\ref{1a}) leads to a simple class of
solutions (\ref{5a})--(\ref{5d}) which are written in terms of elementary functions. As an aside we
observe that $\a$ does not appear in the solutions (\ref{5a})--(\ref{5d}).  Thus the constant of
separability $\a$ can be taken to be zero, and $C$ can be taken to be unity with no loss of
generality.  Indeed as we shall demonstrate later, our  Lie analysis obviates the need for the introduction of $\a$ and $C$.

We will show that, while (\ref{1a}) is an {\it ad hoc} assumption, the reason for its feasibility lies
in the group theoretic properties of  (\ref{don}).  Utilising the full group properties of (\ref{don})
we can provide further solutions to complement (\ref{5a})--(\ref{5d}).

\section{Lie analysis}
The basic feature of  Lie analysis  requires the determination of the
one--parameter ($\ve$) Lie group of transformations
\bq
\bar{t} &=& t + \ve \tau(t,r,u) + O(\ve^2) \n \\
\bar{r} &=& r + \ve \xi(t,r,u) + O(\ve^2) \label{2.2} \\
\bar{u} &=& u + \ve \eta(t,r,u) + O(\ve^2) \n
\eq
that leaves the solution set of a differential equation invariant. (The full details can be found in a number of excellent texts (Bluman and Kumei 1989, Olver 1993).) In order to obtain (\ref{2.2}) we need to determine their  ``generator'' \be  G = \tau \frac{\p\ }{\p t} + \xi \frac{\p\ }{\p
r} + \eta \frac{\p\ }{\p u} \label{2.4} \ee
(also called a symmetry of the differential equation)
which is a set of vector fields.

The determination of these generators  is a straight forward, albeit
tedious process.  Fortunately, a number of computer algebra packages are available to aid the practitioner
(Hereman 1994).  While some modern packages have been developed (Dimas and Tsoubelis 2005, Cheviakov 2007), we have found the  package {\tt PROGRAM LIE}  (Head 1993) to be the most useful in practice. Indeed it is quite remarkable how accomplished such an old package is -- it often outperforms its modern counterparts!

Utilising {\tt PROGRAM LIE}, we can demonstrate that (\ref{don}) admits the following Lie point symmetries/vector fields:
\numparts
\begin{eqnarray}
G_1 &=& \frac{\p\ }{\p V} \label{6a} \\
G_2 &=& \frac{\p \ }{\p r} \label{6b} \\
G_3 &=& t^3 \frac{\p\ }{\p V} \label{6c} \\
G_4 &=& t \frac{\p\ }{\p r} + r \frac{\p\ }{\p V} \label{6d} \\
G_5 &=& t \frac{\p\ }{\p t} - (V+t) \frac{\p\ }{\p V} \label{6e} \\
G_6 &=& r \frac{\p\ }{\p r} + (t+2V)\frac{\p\ }{\p V} \label{6f}
\end{eqnarray}
\endnumparts
with the nonzero Lie bracket relationships
\be
\begin{array}{lcl}
[G_1, G_5] = - G_1 &\qquad& [G_1,G_6]=2G_1 \\
{}[G_2,G_4] = G_1 &\qquad& [G_2,G_6]=G_2  \\
{}[G_3,G_5] = - 4 G_3 &\qquad& [G_3,G_6]=2 G_3  \\
{}[G_4,G_5] = - G_4 & \qquad& [G_4,G_6]=G_4
\end{array} \label{7}
\ee for the given fields. As a result, the symmetries
(\ref{6a})--(\ref{6f}) form a six--dimensional indecomposable
solvable Lie algebra, $L$ (Rand {\it et al}, 1988).  While $L$ is
not nilpotent, its first derived Lie subalgebra
$L^{(1)}=<G_1,G_2,G_3,G_4>$  is nilpotent and also represents the
nilradical of $L$. Further information about such Lie algebras can
be found in (Turkowski, 1990).

\section{Known solutions}

 It is possible to demonstrate that the solutions (\ref{5a})--(\ref{5d}) are a natural consequence of a subset of these symmetries.  If we take the combination
\bq \tilde{G} &=& c_1 G_1 + c_2 G_2 + G_5 \n \\
&=& t \frac{\p\ }{\p t} + (c_1 - t + c_2 t^3 - V)\frac{\p\ }{\p V}
\label{8} \eq the partial differential equation (\ref{don}) is
reduced to the ordinary differential equation \be
UU_{rr}-U_{r}{}^2+U = 0 \label{sh2}\ee as the essential equation
governing the gravitational dynamics where $ V = c_1 - \frac{t}{2} +
\frac{c_2 t^3}{4} + \frac{U}{t}$.

On comparing (\ref{2}) and (\ref{sh2}) we can identify the function $f$ with $U$. We are now in a
position to make a number of comments relating to the underlying assumption (\ref{1a}) in the Gupta and
Sharma (1996) solutions. We observe that the $t$-dependence arises naturally
because of the choice of the symmetry $\tilde{G}$. Thus the temporal dependence is not arbitrary as suggested
 by the function $h(t)$ in the choice (\ref{1a}). It is not necessary to solve any differential equation to obtain
 the form of $h(t)$ given by (\ref{3}).  In addition, the solutions of (\ref{sh2}) are the same
as (\ref{4a})--(\ref{4d}) with $\a = 0$ and $C = 1$.  As stated earlier, the final solutions (\ref{5a})--(\ref{5d}) do
not include these constants. Thus the Lie symmetry $\tilde{G}$ leads directly to the canonical
 form of the solution to (\ref{don}) without the need to introduce spurious arbitrary functions and parameters.

\section{Group invariant solutions}
We now seek to utilise the Lie point symmetries in a systematic manner to generate new solutions.
 These new solutions are termed group invariant solutions as they will be invariant under the group
 generated by the symmetry used to find them. The advantage of using Lie point symmetries is that
 we are guaranteed that the variable combinations obtained will always result in an equation in the
 new variables - no further ``consistency'' conditions are needed. As we are dealing with a $1+1$
 partial differential equation here, we will always be able to find an ordinary differential equation
 in the new variables defined by the symmetries.

\subsection{The optimal system}
Given that equation (\ref{don}) has the six symmetries
(\ref{6a})-(\ref{6f}), we can find group invariant solutions using
each symmetry individually, or any
 linear combination of symmetries. However, taking all possible combinations
  into account is overly excessive.  It turns out (Olver, 1993), that one
  only need consider a subspace of this vector space.  We use
 the subalgebraic structure of the symmetries (\ref{6a})-(\ref{6f})
 of the system (\ref{don}) to construct an  optimal
system of one-dimensional subgroups. Such an optimal system of
subgroups is determined by classifying the orbits of the
infinitesimal adjoint representation of a Lie group on its Lie
algebra obtained by using its infinitesimal generators. All group
invariant solutions can be transformed to those obtained via this
optimal system. The process is algorithmic and can be found in
(Olver 1993).  Here we only summarise the final results.

\begin{table}
\caption{One  symmetry: generators and {\sc odes}.\label{table1}}
\begin{indented} \item[]
\begin{tabular}{@{}lll}
\br
 Generator & Invariants & {\sc ode}  \\
 \mr
$ G_{2}$ & $y = t$ & $yU_{yy}-2U_{y}-1 = 0$   \\
&$V = U(t)$ & \\
 $G_{3}$ & $y = t$ & No {\sc ode} exists   \\
 & $U(t)=r$ &\\
 $G_{4}$ & $y = t$ & $yU_{yy}-2U_{y}-1 =0$  \\
&$V = \frac{r^2}{2t}+U(t)$& \\
 $G_{5}$ & $y = r$ & $ UU_{yy}-(U_{y})^2+U=0$  \\
 &$V = -\frac{t}{2}+\frac{1}{t}U(r)$&  \\
\br
\end{tabular}
\end{indented}
\end{table}

\begin{table}
\caption{One  symmetry: {\sc pde} solutions.\label{table2}}
\begin{indented} \item[]
\begin{tabular}{@{}ll}
\br
Generator & Solution to  {\sc pde} \\
\mr
 $G_{2}$  & $V = A+\frac{1}{3}Bt^3-\frac{t}{2}$\\
$ G_{3}$  & No solution to {\sc pde} \\
 $G_{4}$ & $V = \frac{r^2}{2t}+A+\frac{1}{3}Bt^3-\frac{t}{2}$\\
 $G_{5}$  & $V = -\frac{t}{2}+\frac{1}{t}\left[\frac{1}{2m^2}(\sin (\sqrt{2}mr+m_0)
 + 1)\right] $ \\
 & $V = -\frac{t}{2}$ \\
&$V = -\frac{t}{2}+\frac{1}{t}\left[\frac{1}{2}(r+\beta)^2\right]$ \\
&$V = -\frac{t}{2}+\frac{1}{t}\left[\frac{1}{2m^2}(\cosh
(\sqrt{2}mr+m_0) - 1)\right]$
 \\
\br
\end{tabular}
\end{indented}
\end{table}

In order to obtain group invariant solutions of (\ref{don}) explicitly, the
optimal system yields only the following symmetry combinations
\numparts
\ba
\fl G_{2} = \frac{\p}{\p r}\label{pla}\\
\fl G_{3} = t^3\frac{\p}{\p V}\\
\fl G_{4} = t\frac{\p}{\p r}+r\frac{\p}{\p V}\\
\fl G_{5} = t\frac{\p}{\p t}-(V+t)\frac{\p}{\p V}\\
\fl G_{2}+G_{3} = \frac{\p}{\p r} +t^3\frac{\p}{\p V}\\
\fl G_{2}+G_{4} = (1+t)\frac{\p}{\p r}+r\frac{\p}{\p V}\\
\fl G_{2}+G_{5} = \frac{\p}{\p r}+t\frac{\p}{\p t}-(V+t)\frac{\p}{\p V}\\
\fl G_{3}+G_{4} = (t^3+r)\frac{\p}{\p V}+t\frac{\p}{\p r}\\
\fl aG_{5}+G_{6} = r\frac{\p}{\p r}+(a+1)t\frac{\p}{\p t}+\left[(1-a)V-at\right]\frac{\p}{\p V}\\
\fl G_{2}-G_{3}+G_{4}= (t+1)\frac{\p}{\p r}+(r-t^3)\frac{\p}{\p V}\\
\fl G_{2}+G_{3}+G_{4}= (t+1)\frac{\p}{\p r}+(t^3+r)\frac{\p}{\p V}.
\label{pla1} \ea
\endnumparts
 All solutions of (\ref{don}) which are obtained via other
combinations of point symmetries can be transformed into the
solutions obtained from the combinations above. Here, we have also
taken into account the fact that (\ref{don}) is invariant under the
following involutions:
 $t\rightarrow-t, r\rightarrow-r$ and $V\rightarrow -V$ and so were able to restrict the optimal system further.

 It is clear that the optimal system consists of single elements of the Lie algebra, combinations
  of two elements and combinations of three elements only.  We divide our discussion of the solutions based on this separation.

\begin{table}
\caption{Two  symmetries: generators and {\sc odes}.\label{table3}}
\begin{indented} \item[]
\begin{tabular}{@{}lll}
\br
Generator & Invariants & {\sc ode} \\
\mr $G_{2}+G_{3}$ & $t = y$ & $
yU_{yy}-2U_{y} -4y^6-1= 0$ \\
& $V = rt^3+U(t)$ &
 \\
 $G_{2}+G_{4}$ & $t = y$ & $(1+2y)(-1-2U_{y}+yU_{yy})= 0$ \\
&$V = \frac{r^{2}}{2(1+t)}+U(t) $  & \\
$G_{3}+G_{4}$ & $t = y$ & $
2+y^4+4U_{y}-2yU_{yy}= 0$ \\
& $V = rt^2+\frac{r^{2}}{2t}+U(t)$\\
$G_{2}+G_{5}$ & $r = \ln t + y$ & $U_{yy}+4U+4UU_{yy}+5U_{y}$ \\
 & $V = -\frac{t}{2}+\frac{1}{t}U(r-\ln t)$ & $-4U^{2}_{y}+U_{y}U_{yy}= 0$ \\
$aG_{5}+G_{6}$ & $r = yt^{1/(a+1)}$ & $-(1+2a)^2U^{2}_{y}+y^2U_{yy}$ \\
&$V = -\frac{t}{2}+t^{(1-a)/(1+a)}U(rt^{-1/(1+a)})$ & $+2(-1+a)(1+2a)U(1+U_{yy})$ \\
 & & $ +yU_{y}(2+5a+aU_{yy})= 0$ \\
\br
\end{tabular}
\end{indented}
\end{table}

\begin{table}
\caption{Two symmetries: {\sc pde} solutions.  \label{table4}}
\begin{indented} \item[]\begin{tabular}{@{}ll}
\br
Generator & Solution to  {\sc pde}\\
\mr
$G_{2}+G_{3}$ &  $V = rt^3+A+\frac{1}{3}Bt^3+\frac{t^7}{7}-\frac{t}{2}$\\
 $G_{2}+G_{4}$ &  $V = A\frac{r^2}{2(1+t)}+\frac{1}{3} Bt^3-\frac{t}{2}$ \\
$G_{3}+G_{4}$ & $V = rt^2+\frac{r^2}{2t}+A+\frac{1}{3}Bt^3+\frac{t^5}{20}-\frac{t}{2}$\\
$G_{2}+G_{5}$ & No solution to {\sc pde}\\
$aG_{5}+G_{6}$ & $V =
-\frac{t}{2}+\left(B+\frac{1}{2}\left(-\frac{1}{6}e^{A/2}\left(e^A
-\frac{4r}{t^2}\right)^{3/2}+\frac{e^{A}r}{t^2}\right)\right)t^3$\\
 &  $(a=-\frac{1}{2})$ \\
& $V =
 -\frac{t}{2}-\frac{r^2}{2t}+\frac{C^2r^6}{12t^3}+\frac{\sqrt{C^2r^4-4t^2}}{3Ct}
 -\frac{Cr^4\sqrt{C^2r^4-4t^2}}{12t^3}+D$\\
 & $ (a=1)$ \\
 \br
\end{tabular}
\end{indented}
\end{table}

\subsection{One generator}

In this section we generate solutions to the master equation (\ref{don}) when it admits the individual Lie
 symmetries $G_{2}, G_{3}, G_{4}$ and $G_{5}$. We do not consider the generators $G_{1}$ and $G_{6}$ as
 they do not appear in the optimal system (\ref{pla})--(\ref{pla1}). The procedure of generating the invariants, the
  resultant ordinary differential and finally the solution to the partial differential equation is standard.
  Consequently we do not provide the details of the calculations, instead the relevant results are collated in tabular form.

In table \ref{table1}  we present the {\sc odes} that are generated when a
single Lie symmetry generator is present.  In table \ref{table2} we
give solutions to the partial differential equation (\ref{don}) when
it is invariant under a single generator from the optimal system
(\ref{pla})--(\ref{pla1}). For the generator $G_{3}$ we do not
obtain an invariant involving $V$ and so it is not possible to
generate an ordinary differential equation. Thus there is no
solution possible invariant under  $G_{3}$ alone.
 For the generators $G_{2}$ and $G_{4}$ it is possible to solve the resulting ordinary differential equations and
 obtain explicit forms for the function $V$ given in table \ref{table2}. These are new solutions to equation (\ref{don})
 which have not been obtained previously. For the generator $G_{5}$ we have obtained the ordinary differential equation
\ba UU_{yy}-U_{y}{}^2+U &=& 0 \ea which is of the same form as
(\ref{sh2}) in $\S4$. However it is important to observe that the
characteristics here are different from $\S4$. Consequently the
solutions $V$ generated by the Lie symmetry $G_{5}$, and listed in
table \ref{table2}, comprise a new class of exact solutions to
equation (\ref{don}).

\begin{table}
\caption{Three  symmetries: generators and {\sc
odes}.\label{table5}}
\begin{indented} \item[]
\begin{tabular}{@{}lll}
\br
Generator & Invariants & {\sc ode} \\
\mr $G_{2}+G_{3}+G_{4}$ & $t = y$ & $
 y(5+y(9+y(7+y(2+y^2(2+y)^2))))$ \\
 & $V =
\frac{r^2}{2(1+t)}+\frac{rt^3}{(1+t)}+U(t)$ & $ -(1+y)^3(1+2y)(-2U_{y}+yU_{yy})= -1 $ \\
$G_{2}-G_{3}+G_{4}$ & $t = y$ & $y(5+y(9+y(7+y(2+y^2(2+y)^2))))$ \\
 & $V = \frac{r^2}{2(1+t)}-\frac{rt^3}{(1+t)}+U(t)$& $-(1+y)^3(1+2y)(-2U_{y}+yU_{yy})= -1 $ \\
 \br
\end{tabular}
\end{indented}
\end{table}
\begin{table}
\caption{Three symmetries: {\sc pde} solutions.  \label{table6}}
\begin{indented} \item[]\begin{tabular}{@{}ll}
\br
Generator & Solution to  {\sc pde}\\
\mr $G_{2}+G_{3}+G_{4}$ &  $V =
\frac{r^2}{2(1+t)}+\frac{rt^3}{(1+t)}+A+\frac{1}{8}\left(\frac{3t}{4}-\frac{19t^2}{4}+\frac{t^4}{2}+\frac{2t^5}{5}
+\frac{4}{1+t}\right)$\\
 & $+\frac{1}{8}t^3\left((1+B\frac{8}{3})-\frac{3}{8}(1+8t^3)\log(1+2t)
\right)$ \\
 $G_{2}-G_{3}+G_{4}$ &  $V =
 \frac{r^2}{2(1+t)}+\frac{rt^3}{(1+t)}+A+\frac{1}{8}\left(\frac{3t}{4}-\frac{19t^2}{4}+\frac{t^4}{2}+\frac{2t^5}{5}
+\frac{4}{1+t}\right)$ \\
 & $+\frac{1}{8}t^3\left((1+B\frac{8}{3})-\frac{3}{8}(1+8t^3)\log(1+2t)
\right) $\\
 \br
\end{tabular}
\end{indented}
\end{table}

\subsection{Two generators}
We now consider the combinations of two generators which arise in
the optimal system (\ref{pla})--(\ref{pla1}). Table \ref{table3}
contains the invariants and reduced {\sc ode}.  Table \ref{table4}
lists the analytic solutions of the {\sc pde},
 containing all cases that we were able to obtain explicit general solutions.

For the generator $G_{2}+G_{5}$, we have not been able to find a
solution to the resultant {\sc ode} which is highly nonlinear. For
the generator $aG_{5}+G_{6}$ the resultant {\sc ode} is nonlinear
 and difficult to solve in general. However particular solutions
 can be found for the special parameter values $a=-\frac{1}{2}$ and
$a=1$. It is unlikely that the {\sc ode} will yield closed form
solutions for other values of $a$.

Thus we have generated new analytic solutions to (\ref{don}) via
combinations of two symmetries of (\ref{don})
except for one case corresponding to the generator $G_{2}+G_{5}$.

\subsection{Three generators}
It finally remains to  consider the two combinations of three
generators $G_{2}+G_{3}+ G_{4}$ and $G_{2}-G_{3}+G_{4}$which appear
in the optimal system (\ref{pla})-(\ref{pla1}). In both cases it
 is possible to generate the invariants and reduced {\sc ode}
 which are presented in table $5$. The explicit solutions of the {\sc pde}
 are given in table $6$. Both functions obtained
 in table $6$ are new solutions to equation (\ref{don}).

\section{Physical features}
We briefly describe the behaviour of the thermodynamic variables
corresponding to the spacetime (\ref{dad}). The energy density and
the pressure are given by \numparts \ba
8\pi\rho &=& \frac{1}{t^2P}+\frac{2}{P^2}(V_{tt}V_{rr}-V_{rt}{}^{2})\\
8\pi p &=& \frac{1}{t^2P} \ea \endnumparts respectively, where we
have set \be \fl  P = 1+2 V_{t}+V_{r}{}^{2}. \ee For many applications in
cosmology it is necessary that there exist barotropic equations of
state
 in the form $p = p(\rho)$ (Stephani {\it et al} 2003). We find that for the case of a single generator
 of the optimal system (\ref{pla})--(\ref{pla1}) considered in $\S5.2$ there exists a linear equation
 of state. The relevant equations of state are presented in Table \ref{table7}.

The equations of state $p = \rho$ and $p = \frac{1}{3}\rho$ were identified by Gupta and Sharma (1996)
for their class of solutions. We have demonstrated that their result follows because of the existence of the
symmetry $G_{5}$. The Lie symmetry $G_{2}$ produces a new solution with equation of state $p = \rho$.
The generator $G_{4}$ gives another new solution with the linear equation of state $p = \frac{2\pi}{1+2\pi}\rho$.
Such linear equations of state are of importance in relativistic stellar structures and arise in models of
 quark stars (Komathiraj and Maharaj 2007, Mak and Harko 2004, Sharma and Maharaj 2007,
 Witten 1984). Also, in the modelling of anisotropic relativistic matter in the presence of the
 electromagnetic field for strange stars and matter distributions, we need a linear barotropic
 equation of state (Lobo 2006, Thirukkanesh and Maharaj 2008).

\begin{figure}[htbp]
       \centering
       \includegraphics[width=8.00cm]{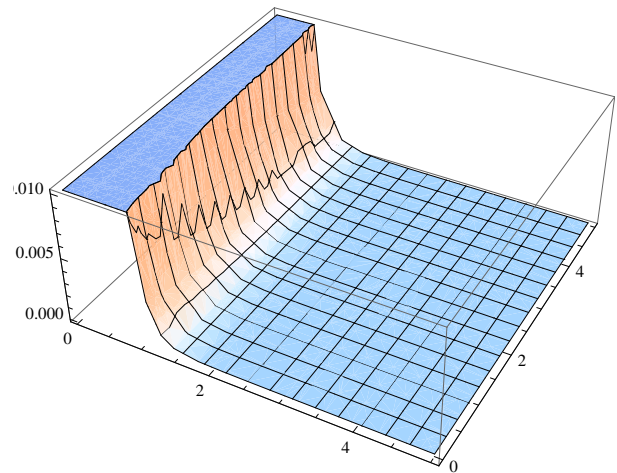}
       \caption{Graph of energy density for $G_{2}+G_{3}$.}
      \label{fig2}
      \end{figure}

\begin{figure}[htbp]
       \centering
       \includegraphics[width=8.00cm]{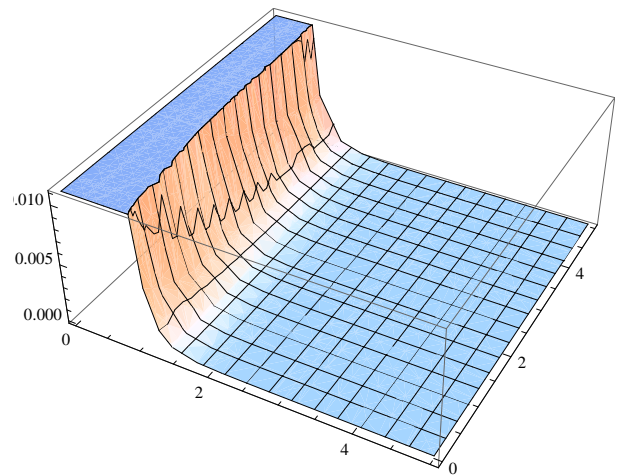}
       \caption{Graph of pressure for $G_{2}+G_{3}$.}
       \label{fig3}
       \end{figure}

For the generators considered in $\S5.3$ and $\S5.4$ there are no
simple barotropic equations of state connecting the energy density
and the pressure. However it is possible to describe the
thermodynamical behaviour graphically. As an example we consider the
solution corresponding to the generator $G_{2}+G_{3}$. The energy
density is \ba \rho &=& \frac{4\pi (B+3r)+3(-3+2\pi)t^4}{16\pi^2
t^4(2B+6r+3t^4)^2}\nonumber \ea and pressure is
 \be \fl p =
\frac{1}{8\pi t^4(2B+6r+3t^4)}.\nonumber \ee Clearly there is no
barotropic equation of state in this case. The behaviour of energy
density has been plotted in figure \ref{fig2} and the pressure is
represented in figure \ref{fig3}. We have generated these plots with the
help of  {\tt Mathematica} (Wolfram 1999). It is clear that there
exist regions of spacetime in which $\rho$ and $p$ are well behaved,
remaining finite, continuous and bounded. It is then viable to study
the behaviour of the thermodynamical quantities such as the
temperature over this region. We point out that plots for $\rho$ and
$p$ for the other combinations of generators in the optimal system
have similar behaviour.

\begin{table}
\caption{Equation of state.\label{table7}}
\begin{indented} \item[]
\begin{tabular}{@{}lll}
\br
 Generator & Solution to {\sc pde} & Equation of state \\
 \mr
 $G_{2}$ & $V = A+\frac{1}{3} Bt^3-\frac{t}{2}$ & $p = \rho$ \\
 $G_{3}$ & No solution to {\sc pde} & No equation of state \\
 $G_{4}$ & $V = \frac{r^2}{2t}+A+\frac{1}{3}B t^3-\frac{t}{2} $ & $p = \frac{2\pi}{1+2\pi} \rho$\\
 $G_{5}$ & $V = -\frac{t}{2}$ & $p = \rho$ \\
& $V = -\frac{t}{2}+\frac{1}{t}\left[\frac{1}{2}(r+\beta)^2\right]$ & $p = \frac{1}{3} \rho$ \\
\br
\end{tabular}
\end{indented}
\end{table}

\section{Conclusion}
A variety of new exact solutions of the governing equation
(\ref{don}), using the Lie method of infinitesimal generators, have
been obtained. Previously known solutions where shown to be
characterized by a particular Lie generator and are contained, as a
special case, in our new family of solutions. We considered each
element in the  optimal system of one-dimensional subgroups and
reduced the master equation to an ordinary differential equation. We
were in a position to solve the resulting equations and obtain
several new solutions for the gravitating model. A pleasing feature
of our analysis is that several models generated  admit a linear
barotropic equation of state. The pivotal equation has six Lie point
symmetries with eight nonzero Lie bracket relationships which
generates the optimal system. It is this geometric structure which
has enabled us to show that equation (\ref{don}) has a rich
structure.

~\\
{\bf Acknowledgements}\\
AMM and KSG thank the National Research Foundation and the
University of KwaZulu-Natal for financial support. SDM acknowledges
that this work is based upon research supported by the South African
Research Chair Initiative of the Department of Science and
Technology and the National Research Foundation. We thank Dr J-C
Ndogmo for helpful discussions on the structure of Lie algebras.

~\\

\section*{References}
\begin{harvard}
\item{}
Bluman G W and Kumei S K 1989 {\it Symmetries and Differential Equations} (New York: Springer--Verlag)
\item{}
Cheviakov A F 2007 {\it Comp. Phys. Comm.} {\bf 176} 48
\item{}
Dahia F and Romero C 2002a {\it J. Math. Phys.}  {\bf 43} 3097
\item{}
Dahia F and Romero C 2002b {\it J. Math. Phys.}  {\bf 43} 5804
\item{}
Dahia F, Romero C and Guz M A S  2008 {\it J. Math. Phys.}  {\bf 49} 112501
\item{}
Dimas S and Tsoubelis D 2005 SYM: A new symmetry-finding package for
Mathematica {\it Proceedings of The 10th International Conference in
 Modern Group Analysis} ed Ibragimov N H, Sophocleous C and Pantelis P A
(Larnaca: University of  Cyprus) 64-70
\item{}
Gupta Y K and Sharma J R 1996  {\it Gen. Relativ. Gravit.} {\bf 28}  1447
\item{}
Head A K 1993 {\it Comp. Phys. Comm.}  {\bf 77}  241
\item{}
Hereman W 1994  {\it Euromath. Bull.}  {\bf 1} 45
\item{}
Komathiraj K and Maharaj S D 2007 {\it Int. J. Mod. Phys.} D  {\bf 16} 1803
\item{}
Krasinski A 1997 {\it Inhomogeneous Cosmological models} (Cambridge: Cambridge University Press)
\item{}
Lobo F S N 2006  {\it Class. Quantum Grav.}  {\bf 23} 1525
\item{}
Mak M K and Harko T 2004
{\it Int. J. Mod. Phys.}  D {\bf 13} 149
\item{}
Olver P J 1993 {\it Applications of Lie Groups to Differential Equations} (New York: Springer-Verlag)
\item{}
Rand D, Winternitz P and Zassenhaus H 1988 {\it Linear Algebra Appl. } {\bf 109} 197
\item{}
Sharma R and Maharaj S D 2007  {\it Mon. Not. R. Astron. Soc.}  {\bf 375} 1265
\item{}
Stephani H, Kramer D, MacCallum M A H, Hoenselaers C and Herlt E 2003 {\it Exact solutions to Einstein's field equations}
 (Cambridge: Cambridge University Press)
\item{}
Stephani H  1967b {\it Commun. Math. Phys.}  {\bf 5} 337
\item{}
Stephani H 1967b {\it Commun. Math. Phys.}  {\bf 4} 137
\item{}
Thirukkanesh S and Maharaj S D 2008 {\it Class. Quantum Grav.}  {\bf 25}  235001
\item{}
Turkowski P 1990 {\it J. Math. Phys.}  {\bf 31} 1344
\item{}
Witten E 1984  {\it Phys. Rev.} D {\bf 30} 272
\item{}
Wolfram S 1999 {\it The Mathematica Book} (Champaign: Wolfram Media)
\end{harvard}
\end{document}